\documentclass[showkeys,superscriptaddress,floatfix,aps,10pt,prd]{revtex4-2}
\usepackage{graphicx,epstopdf}
\usepackage[caption=false]{subfig}
\usepackage{appendix}
\usepackage[T1]{fontenc}
\usepackage{lmodern}
\usepackage[dvipsnames,x11names]{xcolor}
\usepackage[colorlinks=true,linkcolor=ForestGreen,citecolor=ForestGreen,urlcolor=NavyBlue]{hyperref}
\usepackage[sort&compress]{natbib}
\usepackage{morefloats}
\usepackage[pdf]{pstricks}
\usepackage{amsmath}
\usepackage{amssymb}
\usepackage{amsfonts}
\usepackage{rotating}
\usepackage{cancel}
\usepackage{mathtools}
\usepackage{bbm}
\usepackage{dsfont}
\usepackage{bbold}
\usepackage{multirow}
\usepackage{ulem}
\usepackage{physics}
\usepackage{orcidlink}
\usepackage{colortbl}

\def\xslash{x\!\!\!\slash }

\begin{document}

\title{$D \bar D_1(2420)$ and $D^* \bar D^*(2400)$ molecular states: Probing their electromagnetic fingerprints}

\author{Ula\c{s}~\"{O}zdem\orcidlink{0000-0002-1907-2894}}%
\email[]{ulasozdem@aydin.edu.tr }
\affiliation{ Health Services Vocational School of Higher Education, Istanbul Aydin University, Sefakoy-Kucukcekmece, 34295 Istanbul, 
T\"{u}rkiye}

\date{\today}
 
\begin{abstract}
As in previous decades, a comprehensive understanding of the intricate internal configuration of hadrons continues to be a central objective within both experimental and theoretical hadron physics. This pursuit plays a pivotal role in advancing our knowledge of QCD and critically evaluating the robustness and accuracy of the theoretical models developed to date. Furthermore, deciphering the underlying mechanisms of exotic states, both those currently observed and those anticipated in future experiments, remains a pressing and unresolved challenge. Motivated by this, in the present study,  we investigate the electromagnetic properties of the $D \bar D_1(2420)$ and $D^* \bar D^*(2400)$ molecular tetraquark states with quantum numbers $J^{PC} = 1^{--}$, using the QCD light-cone sum rule method. These states are analyzed within a hadronic molecular framework, where their magnetic and quadrupole moments are computed to probe internal structure and geometric deformation. Our results reveal distinct electromagnetic signatures, with the magnetic moments primarily dominated by light-quark contributions, and the quadrupole moments suggesting an oblate charge distribution. The findings are compared with prior studies assuming compact tetraquark configurations, emphasizing the sensitivity of electromagnetic observables to the underlying hadronic structure. This analysis provides critical insights into the nature of exotic hadrons and contributes to the broader understanding of QCD dynamics in the non-perturbative regime.
\end{abstract}

\maketitle

\section{motivation}\label{motivation}

The field of hadron spectroscopy has witnessed remarkable progress during the last twenty years, driven by high-precision experimental data from facilities such as the LHC, Belle, and BESIII, along with significant theoretical progress grounded in Quantum Chromodynamics (QCD). These advances have deepened our understanding of QCD’s non-perturbative regime, especially the mechanisms governing color confinement. A breakthrough in this context has been the observation of hadronic states that do not conform to the conventional quark model, which classifies hadrons strictly as mesons or baryons. These anomalous states are interpreted as multiquark configurations, notably tetraquarks and pentaquarks, challenging traditional hadron classification and offering novel insights into the dynamics of strong interactions and exotic QCD matter. 
The first definitive evidence emerged in 2003 with the Belle Collaboration’s discovery of the X(3872) resonance \cite{Belle:2003nnu}, a candidate tetraquark or hadronic molecule. Since then, numerous exotic candidates with possible four- and five-quark content have been identified by collaborations including LHCb, BESIII, Belle, and CDF, establishing the study of exotic hadrons as a rapidly evolving subfield. In parallel, different theoretical approaches are suggested to elucidate their internal structure. The dominant interpretations for tetraquarks are the hadronic molecular picture—viewing them as loosely bound meson–meson systems—and the compact diquark–antidiquark model, which posits a tightly bound color configuration. However, the inherently non-perturbative nature of low-energy QCD renders a definitive understanding of these states elusive.
A comprehensive survey of recent experimental findings and theoretical developments concerning exotic hadrons can be found in Refs.~\cite{Esposito:2014rxa, Esposito:2016noz, Olsen:2017bmm, Lebed:2016hpi, Nielsen:2009uh, Brambilla:2019esw, Agaev:2020zad, Chen:2016qju, Ali:2017jda, Guo:2017jvc, Liu:2019zoy, Yang:2020atz, Dong:2021juy, Dong:2021bvy, Meng:2022ozq, Wang:2025sic, Chen:2022asf}, which collectively address the classification, interpretation, and open theoretical challenges associated with multiquark states.

Recent experimental investigations have led to the identification of several vector hidden-charm tetraquark states, including Y(4260/4230), Y(4360/4320), Y(4390), Y(4500), Y(4660), and Y(4710/4750/4790) with $J^{PC} = 1^{--}$, collectively known as the $Y$ states, by the BaBar \cite{BaBar:2005hhc, BaBar:2012hpr}, Belle \cite{Belle:2007dxy, Belle:2007umv, Belle:2008xmh, Belle:2014wyt}, CLEO \cite{CLEO:2006tct}, and BESIII \cite{BESIII:2014rja, BESIII:2016adj, BESIII:2016bnd, BESIII:2022joj, BESIII:2023cmv, BESIII:2023wsc, BESIII:2023wqy, BESIII:2024jzg, BESIII:2024ext, BESIII:2023rwv} collaborations.  
Very recently, the $e^+e^-\to \pi^+\pi^- h_c$ cross-section was measured, and three vector hidden-charm tetraquark candidates have been reported by the BESIII collaboration \cite{BESIII:2025bce}. The obtained masses and widths are determined as follows: $M_1=(4223.6_{-3.7-2.9}^{+3.6+2.6})~\mathrm{MeV}$, 
$\Gamma_1=(58.5_{-11.4-6.5}^{+10.8+6.7})~\mathrm{MeV}$,
$M_2=(4327.4_{-18.8-9.3}^{+20.1+10.7})~\mathrm{MeV}$,
$\Gamma_2=(244.1_{-27.1-18.0}^{+34.0+23.9})~\mathrm{MeV}$,
and $M_3=(4467.4_{-5.4-2.7}^{+7.2+3.2})~\mathrm{MeV}$,
$\Gamma_3=(62.8_{-14.4-6.6}^{+19.2+9.8})~\mathrm{MeV}$. The statistical significance of the three Breit-Wigner
assumption over the two Breit-Wigner assumption is greater than $5\sigma$.
However, these observed states do not exhibit a clear and direct correspondence with the conventional meson, which is typically characterized by the quantum numbers \(J^{PC} = 1^{--}\). To further understand the underlying nature of these states, different theoretical approaches have been suggested, and significant research efforts have been directed towards exploring them.   Precise measurement of their properties is essential to unraveling their nature. There exist numerous studies devoted to investigating the properties of these tetraquark candidates, and their details can be found in Refs.~\cite{Esposito:2014rxa, Esposito:2016noz, Olsen:2017bmm, Lebed:2016hpi, Nielsen:2009uh, Brambilla:2019esw, Agaev:2020zad, Chen:2016qju, Ali:2017jda, Guo:2017jvc, Liu:2019zoy, Yang:2020atz, Dong:2021juy, Dong:2021bvy, Meng:2022ozq, Wang:2025sic, Chen:2022asf}. Although we will not discuss these details in the present work, we will briefly highlight some of the studies conducted within the framework of QCD sum rules --- the same method employed in this paper --- in order to provide context.
Several studies, reported in Refs.~\cite{Chen:2010ze, Wang:2013exa, Wang:2016mmg, Wang:2018rfw, Sundu:2018toi, Wang:2021qus}, have examined these states’ spectroscopic properties using QCD sum rules.   
In Ref.~\cite{Chen:2010ze}, a comprehensive set of interpolating currents was constructed for the $Y$ states. The spectroscopic parameters of these states were investigated within the QCD sum rule framework for various quantum numbers, including $J^{PC} = 1^{++}$, $1^{--}$, $1^{-+}$, and $1^{+-}$, considering both $[cq][\bar{c}\bar{q}]$ and $[cs][\bar{c}\bar{s}]$ quark configurations. These analyses were conducted under the assumption that the states possess a diquark–antidiquark internal structure. While some of the predicted mass values showed good agreement with experimentally observed $Y$ candidates, discrepancies were found for others. Additionally, the study also addressed possible decay modes and proposed experimental strategies for the detection of these exotic states.
In Ref.~\cite{Wang:2013exa}, the mass and residue of the $Y(4660)$ state were evaluated using the QCD sum rule approach. The analysis suggested that the $Y(4660)$ resonance is compatible with a diquark–antidiquark configuration, either of the type $c\bar{c}s\bar{s}$ or $c\bar{c}(u\bar{u}+d\bar{d})/\sqrt{2}$, with quantum numbers $J^{PC} = 1^{--}$. Additionally, the study ruled out the possibility of assigning the $Y(4360)$ state to a $c\bar{c}u\bar{d}$ tetraquark structure with quantum numbers $J^{PC} = 1^{\pm -}$. 
In Ref.~\cite{Wang:2016mmg}, a comprehensive QCD sum rule analysis was performed employing multiple current operators to investigate both vector and axial-vector tetraquark states. The derived spectroscopic parameters for the $Y$ states provide compelling evidence for the interpretation of the $Y(4660)$ resonance as a compact $[cq][\bar{c}\bar{q}]$ tetraquark configuration with quantum numbers $J^{PC} = 1^{--}$. Furthermore, the results suggest that the $Y(4260)$ and $Y(4360)$ resonances may exhibit mixed charmonium-tetraquark characteristics, indicating a possible hybrid nature for these states.
In Ref.~\cite{Wang:2018rfw}, the authors constructed two distinct tetraquark current types --- $C \otimes \gamma_\mu C$ and $C\gamma_5 \otimes \gamma_5 \gamma_\mu C$ --- to calculate the mass and residue parameters of $Y$ states. Their analysis led to specific spectroscopic assignments: the $Y(4660)$ and $Y(4630)$ resonances were identified as $c\bar{c}s\bar{s}$ vector tetraquark states of the $C \otimes \gamma_\mu C$ type, while the $Y(4360)$ and $Y(4320)$ were characterized as $c\bar{c}q\bar{q}$ vector tetraquarks of the $C\gamma_5 \otimes \gamma_5 \gamma_\mu C$ type. Notably, the study did not find sufficient evidence to classify $Y(4260)$, $Y(4220)$, and $Y(4390)$ as definitive Y states within this framework. 
In Ref.~\cite{Sundu:2018toi}, the $Y(4660)$ state was investigated within the diquark-antidiquark framework, where it was modeled as a compact $[cs][\bar{c}\bar{s}]$ tetraquark configuration. The study comprehensively evaluated the state's mass, decay constant, and strong decay channels, demonstrating that both the predicted mass and total decay width show remarkable consistency with experimental observations.  
In Ref.~\cite{Wang:2021qus}, scalar, pseudoscalar, vector, axial-vector, and tensor antidiquark currents were constructed to investigate the mass spectrum of Y states within the QCD sum rule framework. The results of this analysis supported the interpretation of the $Y(4360)$, $Y(4390)$, and $Y(4660)$ resonances as $[cq][\bar{c}\bar{q}']$ states, carrying quantum numbers $J^{PC} = 1^{--}$.

Beyond the aforementioned states, there is also the possibility of additional vector hidden-charm tetraquarks, which may possess different quark configurations and internal structures.  
In Refs.~\cite{Ozdem:2022kck, Ozdem:2024txt}, we conducted a thorough investigation into the electromagnetic properties of experimentally reported and theoretically predicted vector hidden-charm states via QCD light-cone sum rules, assuming a compact diquark-antidiquark configuration. 
In the present work, we expand this study by calculating both the magnetic and quadrupole moments of $D \bar D_1(2420)$ and $D^* \bar D^*(2400)$ molecular vector hidden-charm tetraquarks, utilizing the QCD light-cone sum rule methodology~\cite{Chernyak:1990ag, Braun:1988qv, Balitsky:1989ry}, with quantum numbers $J^{PC} = 1^{--}$.  The study of the electromagnetic properties of hadrons provides essential insights into their intrinsic nature, internal structure, and quantum numbers, complementing their spectroscopic features. The electromagnetic characteristics, particularly the magnetic moments, are intrinsically linked to the spatial distribution of quarks and gluons within the hadron. The magnetic moments reflect the distribution of charge and magnetization within the particle. Investigating higher-order multipole moments, including quadrupole moments, offers a promising avenue for advancing our understanding of these exotic states. 
A substantial body of literature exists that focuses on the electromagnetic properties of both hidden-charm and doubly-heavy tetraquark states. These investigations seek to elucidate the internal quark-gluon dynamics and fundamental characteristics of these intricate hadronic structures~\cite{Ozdem:2024dbq, Ozdem:2024lpk, Mutuk:2023oyz, Wang:2023bek, Ozdem:2023rkx, Ozdem:2023frj, Lei:2023ttd, Zhang:2021yul, Azizi:2023gzv, Ozdem:2022eds, Ozdem:2022kck, Xu:2020qtg, Wang:2017dce, Ozdem:2022yhi, Wang:2023vtx, Ozdem:2021hmk, Azizi:2021aib, Ozdem:2021hka, Xu:2020evn, Ozdem:2021yvo, Ozdem:2017exj, Ozdem:2017jqh, Ozdem:2024rrg, Mutuk:2024vzv, Ozdem:2024jmb}. 
While considerable advancements have been achieved, it is evident that additional research is essential to attain a more comprehensive understanding of the electromagnetic characteristics of these states, particularly those with intricate internal configurations.

 This work is organized as follows: Section~\ref{formalism} develops the theoretical foundation using QCD light-cone sum rules, systematically deriving the correlation functions within both hadronic and QCD representations. The formalism yields analytical expressions for both magnetic and quadrupole moments following standard methodology. Section~\ref{numerical} provides a detailed numerical analysis of these electromagnetic moments, with a discussion of the obtained results. Finally, Section~\ref{sum} offers a summary of the findings.
 
 \begin{widetext}
 
\section{Theoretical Foundations}\label{formalism}

In this section, we present the derivation of the QCD light-cone sum rules, which are utilized to compute the magnetic and quadrupole moments of $D \bar D_1(2420)$ and $D^* \bar D^*(2400)$ molecular $Y$ states. To establish the formalism required for this analysis, we begin by considering the following correlation function, which is a crucial component of the method.
\begin{equation}
 \label{edmn01}
\Pi _{\alpha \beta }(p,q)=i\int d^{4}x\,e^{ip\cdot x}\langle 0|\mathcal{T}\{J_{\alpha}(x)
J_{\beta }^{ \dagger }(0)\}|0\rangle_{F}, 
\end{equation}
where $F$ represents the external background electromagnetic field, and $J_{\alpha}(x)$ stands for the interpolating current of the $Y$ states.  For the $D \bar D_1(2420)$ and $D^* \bar D^*(2400)$ molecular tetraquark states, the following interpolating currents can be formulated \cite{Wang:2016wwe}:
\begin{align}
\label{curr1}
J_{\alpha }^{D \bar D_1}(x) &= \frac{i}{\sqrt{2}} \big[\bar u^{a^T} (x) \gamma_5 c^a (x)\big]\big[\bar c^b (x)  \gamma_\alpha \gamma_5  d^{b^T} (x) \big],\\
J_{\alpha }^{D^* \bar D^*}(x) &= \frac{1}{\sqrt{2}} \big[\bar u^{a^T} (x)  c^a (x)\big]\big[\bar c^b (x)  \gamma_\alpha  d^{b^T} (x) \big],
\label{curr4}
\end{align}
where $a$ and $b$ are color indices. 

These interpolating currents possess meson--meson structures and are constructed as local products of scalar--vector currents corresponding to the \( D \) and \( \bar{D}_1 \) (or \( D^* \) and \( \bar{D}^* \)) mesons. This choice of interpolating current is motivated by its meson--meson structure, which aligns with the expected quantum numbers and binding characteristics of hadronic molecules. They are therefore expected to couple effectively to the molecular configurations of the \( D\bar{D}_1 \) and \( D^*\bar{D}^* \) systems. 
 However, it is worth noting that the interpolating currents mentioned above couple not only to molecular states but also to diquark–antidiquark (compact tetraquark) configurations. This is because, through a Fierz transformation, a molecular-type current can be decomposed into a sum of diquark–antidiquark-type currents with certain numerical coefficients \cite{Wang:2020rcx}. In other words, a molecular current can be represented as a specific linear combination of compact-type currents. Conversely, a compact tetraquark current can also be re-expressed in terms of molecular-type structures, as shown in \cite{Chen:2022sbf}. However, a notable distinction exists between an individual molecular-type current and an individual diquark–antidiquark-type current. For instance, each compact tetraquark current is typically a linear combination of multiple independent molecular currents. A well-known example illustrating this behavior is the light scalar-isoscalar sigma meson, for which the compact tetraquark current, or a suitable mixing of currents, provides a better phenomenological description than the conventional pion–pion molecular current. Distinguishing between compact tetraquark and molecular structures remains one of the central aims of our systematic investigation, driven by the need to understand the underlying nature of these exotic states.

By utilizing these interpolating currents, we will investigate the electromagnetic properties of these molecular states using the QCD light-cone sum rules method.  
The procedure to apply the QCD light-cone sum rules is outlined as follows:

\begin{itemize}
\item The first step involves expressing the correlation function in terms of hadronic parameters, such as mass, magnetic moment, and form factors. This approach is termed the ``hadronic formulation.''
    
\item The second step involves expressing the same correlation function in terms of quantities associated with the quark-gluon degrees of freedom and distribution amplitudes. This is known as the ``QCD formulation.''

\item Finally, the two formulations are reconciled based on the premise of quark-hadron duality. To suppress unwanted contributions, a double Borel transformation along with continuum subtraction is applied. This process leads to the derivation of the sum rules for the physical parameters of interest.
\end{itemize}

\subsection{Hadronic formulation}

To obtain the hadronic formulation of the correlation function, we begin by inserting a complete set of intermediate states corresponding to the $Y$ states with identical quantum numbers as the interpolating currents. This is done within the framework of the correlation function, and an integration over the spacetime variable \( x \) is then performed. The expression that results from this procedure is:
\begin{align}
\label{edmn04}
\Pi_{\alpha\beta}^{Had}(p,q) &= \frac{\langle 0 \mid J_\alpha(x) \mid Y(p, \varepsilon^i) \rangle}{p^2 - m_{Y}^2} \langle Y(p, \varepsilon^i) \mid Y(p+q, \varepsilon^f) \rangle_F \frac{\langle Y(p+q, \varepsilon^f) \mid J_\beta^\dagger(0) \mid 0 \rangle}{(p+q)^2 - m_{Y}^2} \nonumber\\
&+ \mbox{continuum and higher states}.
\end{align}

From the expression above, it is evident that the matrix elements involved need to be explicitly defined. These elements are presented as follows \cite{Brodsky:1992px}:
\begin{align}
\label{edmn05}
\langle 0 \mid J_\alpha (x) \mid Y (p, \varepsilon^i) \rangle &=  \lambda_{Y} \varepsilon_\alpha^i\,,\\
\langle Y (p+q, \varepsilon^{f}) \mid J_{\beta }^{\dagger } (0) \mid 0 \rangle &= \lambda_{Y} \varepsilon_\beta^{* f}\,,\\
\langle Y(p,\varepsilon^i) \mid  Y (p+q,\varepsilon^{f})\rangle_F &= - \varepsilon^\gamma (\varepsilon^{i})^\mu (\varepsilon^{f})^\nu
\Big[ G_1(Q^2)~ (2p+q)_\gamma ~g_{\mu\nu}  
+ G_2(Q^2)~ ( g_{\gamma\nu}~ q_\mu -  g_{\gamma\mu}~ q_\nu)
\nonumber\\ 
&
- \frac{1}{2 m_{Y}^2} G_3(Q^2)~ (2p+q)_\gamma
q_\mu q_\nu  \Big],\label{edmn06}
\end{align}
where \( \lambda_{Y} \) represents the current coupling, \( \varepsilon_\alpha^{\mathrm{i}} \)(\( \varepsilon_\beta^{*\mathrm{f}} \))  denotes the polarization vector of the initial(final) \( Y \) tetraquarks, \( \varepsilon^\gamma \) stands for the photon polarization vector, and \( G_i(Q^2) \) correspond to the electromagnetic form factors with \( Q^2 = -q^2 \) being the momentum transfer squared.

By utilizing the expressions above, the hadronic formulation of the correlation function is obtained as follows:
\begin{align}
\label{edmn09}
 \Pi_{\alpha\beta}^{Had}(p,q) &=  \frac{\varepsilon_\rho \, \lambda_{Y}^2}{ [m_{Y}^2 - (p+q)^2][m_{Y}^2 - p^2]}
 \Bigg\{G_1(Q^2)(2p+q)_\rho\Bigg[g_{\alpha\beta}-\frac{p_\alpha p_\beta}{m_{Y}^2}
 -\frac{(p+q)_\alpha (p+q)_\beta}{m_{Y}^2}+\frac{(p+q)_\alpha p_\beta}{2m_{Y}^4}\nonumber\\
 & \times (Q^2+2m_{Y}^2)
 \Bigg]
 + G_2 (Q^2) \Bigg[q_\alpha g_{\rho\beta}  
 - q_\beta g_{\rho\alpha}-
\frac{p_\beta}{m_{Y}^2}  \big(q_\alpha p_\rho - \frac{1}{2}
Q^2 g_{\alpha\rho}\big) 
+
\frac{(p+q)_\alpha}{m_{Y}^2}  \big(q_\beta (p+q)_\rho+ \frac{1}{2}
Q^2 g_{\beta\rho}\big) 
\nonumber\\
&-  
\frac{(p+q)_\alpha p_\beta p_\rho}{m_{Y}^4} \, Q^2
\Bigg]
-\frac{G_3(Q^2)}{m_{Y}^2}(2p+q)_\rho \Bigg[
q_\alpha q_\beta -\frac{p_\alpha q_\beta}{2 m_{Y}^2} Q^2 
+\frac{(p+q)_\alpha q_\beta}{2 m_{Y}^2} Q^2
-\frac{(p+q)_\alpha q_\beta}{4 m_{Y}^4} Q^4\Bigg]
\Bigg\}\,.
\end{align}

Rather than the form factors $G_1(Q^2)$, $G_2(Q^2)$, and $G_3(Q^2)$ defined above, the magnetic $F_M(Q^2)$ and quadrupole $F_{\mathcal{D}}(Q^2)$ form factors are experimentally more accessible, and their explicit forms are defined in terms of the form factors $G_1(Q^2)$, $G_2(Q^2)$, and $G_3(Q^2)$ as follows:
\begin{align}
\label{edmn07}
&F_M(Q^2) = G_2(Q^2)\,,\nonumber \\
&F_{\cal D}(Q^2) = G_1(Q^2)-G_2(Q^2)+(1+\lambda) G_3(Q^2)\,,
\end{align}
where $\lambda=Q^2/4 m_{Y}^2$ is a  kinematic factor  with $Q^2=-q^2$.
In standard contexts, the equation above is sufficient to obtain the electromagnetic properties of the 
Y states. However, since our analysis is focused on real photon interactions, we need to derive the expressions at the point \( Q^2 = 0 \).  At zero momentum transfer, i.e., in the static limit, these form factors are proportional to the usual static quantities: the magnetic ($\mu$) and quadrupole ($\mathcal{D}$) moments. The corresponding equations are given as follows:
\begin{align}
\label{edmn08}
&e F_M(0) = 2 m_{Y} \mu_Y \,, \nonumber\\
&e F_{\cal D}(0) = m_{Y}^2 {\cal D}_Y\,.
\end{align}

Thus, the necessary formulas for the magnetic and quadrupole moments have been derived, thereby completing the hadronic formulation of the correlation function. The next step involves analyzing the correlation function in terms of the quark-gluon interactions and their associated characteristics.

 \subsection{QCD formulation}

Within the QCD formulation for the correlation function, the corresponding interpolating currents for the $D\bar D_1$ and $D^\ast\bar D^\ast$ states are integrated into the correlation function as presented in Eq. (\ref{edmn01}). Following this, Wick's theorem is used to carry out all required contractions and to obtain the corresponding equations. Following these steps, the QCD formulation of the correlation function for the $D\bar D_1$ and $D^\ast\bar D^\ast$ states is derived as follows:
%
\begin{align}
\label{eq:QCDSide}
\Pi _{\alpha \beta }^{\mathrm{QCD}- D\bar D_1}(p,q)&=-\frac{i}{2}\int d^{4}xe^{ip\cdot x}   \langle 0 \mid  
\Bigg\{\mathrm{Tr}\Big[ \gamma _{5}{S}_{c}^{aa^{\prime }}(x)\gamma _{5} S_{u}^{a^{\prime }a}(-x)\Big]    
\mathrm{Tr}\Big[  \gamma _{\alpha } \gamma _{5} S_{d}^{bb^{\prime}}(x) \gamma _{5} \gamma _{\beta }  S_{c}^{b^{\prime }b}(-x)\Big] 
\nonumber\\
&-\mathrm{Tr}\Big[ \gamma _{5}{S}_{c}^{aa^{\prime }}(x)\gamma _{5} \gamma_\beta S_{u}^{a^{\prime }a}(-x)\Big]    
\mathrm{Tr}\Big[  \gamma _{\alpha } \gamma _{5} S_{d}^{bb^{\prime}}(x) \gamma _{5}   S_{c}^{b^{\prime }b}(-x)\Big] 
\nonumber\\
&-\mathrm{Tr}\Big[ \gamma_\alpha \gamma _{5}{S}_{c}^{aa^{\prime }}(x)\gamma _{5} S_{u}^{a^{\prime }a}(-x)\Big]    
\mathrm{Tr}\Big[ \gamma _{5} S_{d}^{bb^{\prime}}(x) \gamma _{5} \gamma _{\beta }  S_{c}^{b^{\prime }b}(-x)\Big] 
\nonumber\\
&+\mathrm{Tr}\Big[ \gamma _{\alpha }  \gamma _{5}{S}_{c}^{aa^{\prime }}(x)\gamma _{5} \gamma _{\beta } S_{u}^{a^{\prime }a}(-x)\Big]    
\mathrm{Tr}\Big[  \gamma _{5} S_{d}^{bb^{\prime}}(x) \gamma _{5}   S_{c}^{b^{\prime }b}(-x)\Big] 
\Bigg\}\mid 0 \rangle_{F} ,  
\end{align}
\begin{align}
\label{eq:QCDSide1}
\Pi _{\alpha \beta }^{\mathrm{QCD}- D^*\bar D^*}(p,q)&=\frac{1}{2}\int d^{4}xe^{ip\cdot x}   \langle 0 \mid  
\Bigg\{\mathrm{Tr}\Big[ {S}_{c}^{aa^{\prime }}(x) S_{u}^{a^{\prime }a}(-x)\Big]    
\mathrm{Tr}\Big[  \gamma _{\alpha }  S_{d}^{bb^{\prime}}(x)  \gamma _{\beta }  S_{c}^{b^{\prime }b}(-x)\Big] 
\nonumber\\
&-\mathrm{Tr}\Big[ {S}_{c}^{aa^{\prime }}(x) \gamma_\beta S_{u}^{a^{\prime }a}(-x)\Big]    
\mathrm{Tr}\Big[  \gamma _{\alpha } S_{d}^{bb^{\prime}}(x)   S_{c}^{b^{\prime }b}(-x)\Big] 
\nonumber\\
&-\mathrm{Tr}\Big[ \gamma_\alpha {S}_{c}^{aa^{\prime }}(x) S_{u}^{a^{\prime }a}(-x)\Big]    
\mathrm{Tr}\Big[ S_{d}^{bb^{\prime}}(x)  \gamma _{\beta }  S_{c}^{b^{\prime }b}(-x)\Big] 
\nonumber\\
&+\mathrm{Tr}\Big[ \gamma _{\alpha }  {S}_{c}^{aa^{\prime }}(x) \gamma _{\beta } S_{u}^{a^{\prime }a}(-x)\Big]    
\mathrm{Tr}\Big[  S_{d}^{bb^{\prime}}(x)    S_{c}^{b^{\prime }b}(-x)\Big] 
\Bigg\}\mid 0 \rangle_{F}.
\end{align}

The propagator structures for both charm and light quark fields adopt the following representation~\cite{Balitsky:1987bk, Belyaev:1985wza}:
\begin{align}
\label{edmn13}
S_{q}(x)&= S_q^{free}(x) 
-\frac {i g_s }{16 \pi^2 x^2} \int_0^1 du \, G^{\mu \nu} (ux)
\bigg[\bar u \rlap/{x} 
\sigma_{\mu \nu} + u \sigma_{\mu \nu} \rlap/{x}
 \bigg],\\
%
S_{Q}(x)&=S_Q^{free}(x)
-i\frac{m_{Q}\,g_{s} }{16\pi ^{2}}  \int_0^1 du \,G^{\mu \nu}(ux)\bigg[ (\sigma _{\mu \nu }{\xslash}
+{\xslash}\sigma _{\mu \nu }) 
    \frac{K_{1}\big( m_{Q}\sqrt{-x^{2}}\big) }{\sqrt{-x^{2}}}
 +2\sigma_{\mu \nu }K_{0}\big( m_{Q}\sqrt{-x^{2}}\big)\bigg],
 \label{edmn14}
\end{align}%
with  
\begin{align}
 S_q^{free}(x)&=\frac{1}{2 \pi x^2}\Big(i \frac{\xslash}{x^2}- \frac{m_q}{2}\Big),\\
 S_c^{free}(x)&=\frac{m_{c}^{2}}{4 \pi^{2}} \Bigg[ \frac{K_{1}\big(m_{c}\sqrt{-x^{2}}\big) }{\sqrt{-x^{2}}}
+i\frac{{\xslash}~K_{2}\big( m_{c}\sqrt{-x^{2}}\big)}
{(\sqrt{-x^{2}})^{2}}\Bigg],
\end{align}
where $G^{\mu\nu}$ denotes the gluon field-strength tensor, and $K_i$'s represent modified Bessel functions of the second kind.

There are two classes of contributions in Eqs.~(\ref{eq:QCDSide})-(\ref{eq:QCDSide1}), distinguished by the manner in which the photon interacts with the quark lines. In the first case, the photon interacts perturbatively with the quark line via the standard QED interaction, commonly referred to as the short-distance interaction. The second class involves the non-perturbative interaction of the photon with quarks, which is described by the photon light-cone distribution amplitude, known as the long-distance interaction. 
A complete theoretical treatment necessitates the inclusion of both contributions to ensure robust and self-consistent results. Our implementation strategy for these effects proceeds as follows:
\begin{itemize}
 \item The following formulation will be sufficient when including perturbative contributions in the analysis.
\begin{align}
\label{free}
S_{c(q)}^{free}(x) \longrightarrow \int d^4z\, S_{c(q)}^{free} (x-z)\,\rlap/{\!A}(z)\, S_{c(q)}^{free} (z)\,.
\end{align}
In this approach, one of the light or heavy quark propagators is substituted in the equation above, while the free part of the remaining propagators is retained.   This procedure corresponds to setting $\bar T_4^{\gamma} (\underline{\alpha}) = 0$ and $S_{\gamma} (\underline {\alpha}) = \delta(\alpha_{\bar q})\delta(\alpha_{q})$, in line with the treatment of three-particle light-cone distribution amplitudes, as detailed in Ref. \cite{Li:2020rcg}. 

\item The inclusion of non-perturbative effects in the current framework requires implementation through the following mathematical formulation:
 \begin{align}
\label{edmn21}
S_{q,\alpha\beta}^{ab}(x) \longrightarrow -\frac{1}{4} \big[\bar{q}^a(x) \Gamma_i q^b(0)\big]\big(\Gamma_i\big)_{\alpha\beta},
\end{align}
where $\Gamma_i = \{\textbf{1}$, $\gamma_5$, $\gamma_\alpha$, $i\gamma_5 \gamma_\alpha$, $\sigma_{\alpha\beta}/2\}$.  Within this framework, the calculation involves replacing a single quark propagator in the aforementioned equation while treating the remaining propagators in their full form. 
The incorporation of non-perturbative effects in the analytical framework generates specific matrix elements, including $\langle \gamma(q)|\bar{q}(x) \Gamma_i G_{\alpha\beta}q(0)|0\rangle$ and $\langle \gamma(q)|\bar{q}(x) \Gamma_i q(0)|0\rangle$. These matrix elements are fundamentally characterized through photon distribution amplitudes (DAs). Photon DAs encapsulate non-perturbative QCD effects and are essential for evaluating the correlation function. Their explicit forms, given in Ref. \cite{Ball:2002ps}, include contributions up to twist-4 accuracy. It should be emphasized that the photon DAs employed in this study account solely for the contributions of light quarks. Although, in principle, a photon may also be radiated at long distances by a charm quark, such effects are negligible in our analysis. Technically, the matrix elements of nonlocal operators are expressed in terms of DAs, quark condensates, and certain non-perturbative constants. Since the contributions of these constants are already negligible for light quarks, their impact becomes even less significant for charm quarks. Notably, charm quark condensates scale inversely with the heavy-quark mass, i.e., $\sim 1/m_c$, and are therefore strongly suppressed~\cite{Antonov:2012ud}. As a result, we exclude long-distance contributions involving charm quark DAs from our calculations and retain only the short-distance photon emission from charm quarks, as formulated in Eq.~(\ref{free}). 
 The analytical framework for systematically combining both perturbative and non-perturbative components is comprehensively presented in Refs.~\cite{Ozdem:2022vip,OZDEM:2024jlw, Ozdem:2022eds}.  By incorporating these theoretical refinements — specifically the inclusion of both perturbative and non-perturbative contributions — the complete QCD formulation of the correlation function is obtained.

\end{itemize}

\subsection{Sum rules for magnetic and quadrupole moments}

Finally, by equating the coefficients of the unique Lorentz structure $(\varepsilon \cdot p)(q_\alpha p_\beta - p_\alpha q_\beta)$ for the magnetic moment and $(\varepsilon \cdot p) q_\alpha q_\beta $ for the quadrupole moment in both the QCD and hadronic formulations, we deduce the QCD light-cone sum rules. These sum rules enable the extraction of the magnetic and quadrupole moments of the corresponding molecular tetraquark states, expressed in terms of QCD parameters, hadronic inputs, and the auxiliary parameters $\rm{s_0}$ and $\rm{M^2}$. The resulting predictions are presented below: 
\begin{align}
\label{jmu1}
 &\mu_{D \bar D_1} \, e^{-\frac{m_{{D \bar D_1}}^{2}}{\rm{M^2}}}   = \frac{m_{{D \bar D_1}}^{2}}{\lambda^{2}_{{D \bar D_1}} }\,\, \rho_1(\rm{M^2},\rm{s_0}),~~~~~~~~~~
  \mathcal{D}_{D \bar D_1} \, e^{-\frac{m_{{D \bar D_1}}^{2}}{\rm{M^2}}}  = \frac{m_{{D \bar D_1}}^{2}}{\lambda^{2}_{{D \bar D_1}} } \,\, \rho_2(\rm{M^2},\rm{s_0}),  \\
 \nonumber\\
 & \mu_{D^* \bar D^*} \,e^{-\frac{m_{{D^* \bar D^*}}^{2}}{\rm{M^2}}} = \frac{m_{{D^* \bar D^*}}^{2}}{\lambda^{2}_{{D^* \bar D^*}}} \,\, \rho_3(\rm{M^2},\rm{s_0}), ~~~~~
 \mathcal{D}_{D^* \bar D^*} \,e^{-\frac{m_{{D^* \bar D^*}}^{2}}{\rm{M^2}}}  =\frac{m_{{D^* \bar D^*}}^{2}}{\lambda^{2}_{{D^* \bar D^*}}} \,\, \rho_4(\rm{M^2},\rm{s_0}).\label{jmu4}
 \end{align}
Since the expressions $ \rho_i(\rm{M^2},\rm{s_0}) $ share a similar structure, only two of them—$ \rho_1(\rm{M^2},\rm{s_0})$ and $\rho_3(\rm{M^2},\rm{s_0}) $—are explicitly provided below as illustrative examples.
\begin{align}
 \rho_1(\rm{M^2},\rm{s_0})&= \frac{1}{2 ^{24} \times 5^2 \times 7 \pi^5} (e_u-e_d )
 \Big[  387 \, I[0, 5]  \Big]
    \nonumber\\
    &+\frac{  m_c  \langle g_s^2 G^2\rangle \langle \bar q q \rangle }{2 ^{22} \times 3^3 \times 5 \pi^3} (e_u-e_d)\Big[  
   ( 2  I_ 3[\mathcal S] +24 I_ 3[\mathcal T_ 1] + I_ 3[\mathcal T_ 2] - 
       23 I_ 3[\mathcal T_ 4] + 22 I_ 3[\mathcal {\tilde S}]\big) I[0, 1 - 
 648 \chi  I_ 6[\varphi_\gamma] I[0, 2] \Big]
    \nonumber\\
    & +\frac{   \langle g_s^2 G^2\rangle f_{3 \gamma} }{2 ^{25} \times 3^4 \pi^3} (e_u - e_d)\Big[ 23040 m_c^2 I_ 6[\psi_ {\gamma}^{\nu}] I[0, 1] - 
 11 (I_ 2[\mathcal A] + 2 I_ 3[\mathcal A]) I[0, 2] \Big]
    \nonumber\\
    &+\frac{ m_c \langle \bar q q \rangle }{2 ^{19} \times 3^2 \times 5 \times 7 \pi^3} (e_u - e_d) \Big[  63 \Big (-8  (I_ 3[\mathcal T_ 1] - I_ 3[\mathcal T_ 4] + 
          I_ 3[\mathcal {\tilde S}]) - 
    7  I_ 6[\mathbb A]\Big) I[0, 3]  - 
 351 \chi I_ 6[\varphi_\gamma] I[0, 4]\Big]
      \nonumber\\
    &-\frac{f_{3\gamma}}{2 ^{21} \times 3^2 \times 5 \times 7\pi^3} (e_u - e_d)
    \Big[ 1134  I_ 6[\psi^a] I[0, 
   4] -   (3584 m_c^2 I[0, 3] + 
    1377 I[0, 4]) I_ 6[\psi_ {\gamma}^{\nu}] \Big],\\
 \nonumber\\
 %
 \rho_3
 (\rm{M^2},\rm{s_0})&= -\frac{1}{2 ^{23} \times 5^2 \times 7 \pi^5} (e_u-e_d )
 \Big[  387 \, I[0, 5]  \Big]
    \nonumber\\
    &+\frac{  m_c  \langle g_s^2 G^2\rangle \langle \bar q q \rangle }{2 ^{20} \times 3^3 \times 5 \pi^3} (e_u-e_d)\Big[  
   5(  I_ 3[\mathcal S] + 11 I_ 3[\mathcal {\tilde S}]\big) I[0, 1 - 
 324 \chi  I_ 6[\varphi_\gamma] I[0, 2] \Big]
    \nonumber\\
    &-\frac{ m_c \langle \bar q q \rangle }{2 ^{18} \times 3^2 \times 5 \times 7 \pi^3} (e_u - e_d) \Big[  -7 \Big (16  I_ 3[\mathcal {\tilde S}] + 
    9  I_ 6[\mathbb A])\Big) I[0, 3]  + 
 351 \chi I_ 6[\varphi_\gamma] I[0, 4]\Big]
      \nonumber\\
    &+\frac{9 f_{3\gamma}}{2 ^{19} \times  5 \times 7\pi^3} (e_u - e_d)
    \Big[ I_ 6[\psi^a] I[0, 
   4] \Big],
 \end{align}

\noindent where $\langle g_s^2 G^2 \rangle$ represents the gluon condensate, and $\langle \bar{q} q \rangle$ corresponds to the light-quark condensate.
 The functions $I[n,m]$ and $I_i[\mathcal{F}]$ are given by:
\begin{align}
 I[n,m]&= \int_{\mathcal M}^{\rm{s_0}} ds ~ e^{-s/\rm{M^2}}~
 s^n\,(s-\mathcal M)^m,\nonumber\\
  I_2[\mathcal{F}]&=\int D_{\alpha_i} \int_0^1 dv~ \mathcal{F}(\alpha_{\bar q},\alpha_q,\alpha_g) \delta'(\alpha_{\bar q}+ v \alpha_g-u_0),\nonumber\\
   I_3[\mathcal{F}]&=\int D_{\alpha_i} \int_0^1 dv~ \mathcal{F}(\alpha_{\bar q},\alpha_q,\alpha_g)
 \delta(\alpha_ q +\bar v \alpha_g-u_0),\nonumber\\
 I_6[\mathcal{F}]&=\int_0^1 du~ \mathcal{F}(u),\nonumber
 \end{align}
 where  $\mathcal M = 4m_c^2$ and $\mathcal{F}$ refers the designated  photon DAs. 
 
 It is worth noting that the Borel transformations are carried out according to the following standard expressions
\begin{align}
 \mathcal{B}\bigg\{ \frac{1}{\big[ [p^2-m^2_i][(p+q)^2-m_f^2] \big]}\bigg\} \rightarrow e^{-m_i^2/M_1^2-m_f^2/M_2^2}
\end{align}
in the hadronic formulation, and 
\begin{align}
 \mathcal{B}\bigg\{ \frac{1}{\big(m^2- \bar u p^2-u(p+q)^2\big)^{\alpha}}\bigg\} \rightarrow (M^2)^{(2-\alpha)} \delta (u-u_0)e^{-m^2/M^2},
\end{align}
 in the QCD formulation,  where we use
\begin{align*}
 {M^2}= \frac{M_1^2 M_2^2}{M_1^2+M_2^2}, ~~~
 u_0= \frac{M_1^2}{M_1^2+M_2^2}.
\end{align*}
Here, $M_1^2$ and $M_2^2$ denote the Borel parameters associated with the initial and final states, respectively. Given that the same hadronic state appears in both channels, it is reasonable to set $M_1^2 = M_2^2 = 2M^2$ and $u_0 = 1/2$. This approximation ensures a balanced treatment of both sides of the correlation function and is adequate to effectively suppress contributions from higher resonances and the continuum.

It is essential to emphasize that the derivation of Eq. (\ref{edmn04}) is based on the assumption that the hadronic formulation of the QCD light-cone sum rules can be sufficiently estimated using a single pole approach. When considering tetraquark or pentaquark states, it is crucial to substantiate this approximation by introducing additional considerations. This is because the hadronic formulation of the QCD sum rules may involve contributions from intermediate two-meson states~\cite{Weinberg:2013cfa, Lucha:2021mwx, Kondo:2004cr, Lucha:2019pmp}, which are considered unwanted contributions. Therefore, when extracting the parameters for tetraquark or pentaquark states, one should account for unwanted contributions arising from two-meson intermediate states. These unwanted contaminations can either be removed from the QCD sum rules or embedded directly into the pole term parameters. The first method was applied in studies of pentaquarks~\cite{Sarac:2005fn, Wang:2019hyc, Lee:2004xk}, while the second was used to derive tetraquark parameters~\cite{Wang:2015nwa, Agaev:2018vag, Sundu:2018nxt, Albuquerque:2021tqd, Albuquerque:2020hio, Wang:2020iqt, Wang:2019igl, Wang:2020cme}. 
In the present work, it is important that the quark propagator be revised according to the following relation:
\begin{align}
\frac{1}{m^{2}-p^{2}} \longrightarrow  \frac{1}{m^{2}-p^{2}-i\sqrt{p^{2}}\,\Gamma(p)}, \label{eq:Modif}
\end{align}
where $\Gamma(p)$ represents the finite decay width of the multiquark states caused by the intermediate two-meson effects. Incorporating these effects into the sum rules, it has been shown that they influence physical quantities by an amount of roughly $(5-7)\%$~\cite{Wang:2015nwa, Agaev:2018vag, Sundu:2018nxt, Albuquerque:2021tqd, Albuquerque:2020hio, Wang:2020iqt, Wang:2019igl, Wang:2020cme}, a contribution that is smaller than the inherent errors in the sum rule computations. As a result, one can reasonably expect that the electromagnetic characteristics of the tetraquarks and pentaquarks are unaffected by these two-meson intermediate state contributions. Therefore, it is justified to disregard the influence of these intermediate states in the hadronic formulation of the correlation function and to adopt the zero-width single-pole model instead. 
Finally, mathematical formulations for the magnetic and quadrupole moments of the Y tetraquarks have been obtained.

\end{widetext}

\section{Numerical Evaluations}\label{numerical}

The evaluation of magnetic and quadrupole moments via QCD light-cone sum rules necessitates several input parameters. In this analysis, the following values are employed:  $m_u = 2.16 \pm 0.07$ MeV, $m_d = 4.70 \pm 0.07$ MeV, $m_c = 1.27 \pm 0.02\,\mbox{GeV}$~\cite{ParticleDataGroup:2022pth},  $m_{D \bar D_1}=4.36 \pm 0.08$ GeV~\cite{Wang:2016wwe}, $m_{D^* \bar D^*}= 4.78 \pm 0.07$ GeV~\cite{Wang:2016wwe},
  $\lambda_{D \bar D_1}=(3.97 \pm 0.54) \times 10^{-2}$ GeV$^5$~\cite{Wang:2016wwe}, 
  $\lambda_{D^* \bar D^*}= (7.56 \pm 0.84) \times 10^{-2}$ GeV$^5$~\cite{Wang:2016wwe}, $\langle \bar uu\rangle = 
\langle \bar dd\rangle=(-0.24 \pm 0.01)^3\,\mbox{GeV}^3$, 
$\chi= -2.85 \pm 0.5 $ GeV$^{-2}$ \cite{Rohrwild:2007yt}, $f_{3\gamma}=-(0.0039 \pm 0.0020)~$GeV$^2$~\cite{Ball:2002ps}, and $\langle g_s^2G^2\rangle = 0.48 \pm 0.14~ \mbox{GeV}^4$~\cite{Narison:2018nbv}. Photon DAs are among the key input parameters in our numerical calculations. The explicit forms of these DAs and other input parameters are provided in Ref.~\cite{Ball:2002ps}.

As evidenced by Eqs.~(\ref{jmu1})--(\ref{jmu4}), two auxiliary parameters, 
$\rm{s_0}$ (continuum threshold) and $\rm{M^2}$ (Borel mass), play a critical 
role in our calculations. To maintain the robustness and self-consistency of 
our results, it is necessary to establish an optimal range---referred to as 
the working region---within which the computed magnetic and quadrupole moments remain stable under small variations of these parameters. The parameter $\mathrm{s_0}$ serves as a physical scale rather than an arbitrary quantity, marking the threshold beyond which continuum and excited states contribute significantly to the correlation function. While diverse approaches for determining its working region exist in the literature, empirical evidence consistently supports the range: 
$ (m_{Y} + 0.5)^2~\mathrm{GeV}^2 \leq \mathrm{s_0} \leq (m_{Y} + 0.7)^2~\mathrm{GeV}^2.$ This interval has been adopted in our analysis due to its demonstrated reliability in similar theoretical frameworks. 
The working region for $\rm{M^2}$ is constrained by two fundamental criteria: pole dominance (PC) and OPE convergence (CVG). PC ensures that the ground-state contribution dominates over higher-order effects, while CVG guarantees the validity of the truncated operator product expansion. 
These criteria are mathematically implemented as follows:
\begin{align}
 \mbox{PC} &=\frac{\rho_i (\rm{M^2},\rm{s_0})}{\rho_i (\rm{M^2},\infty)} \geq 30 \%,
\\
 \mbox{CVG} (\rm{M^2}, \rm{s_0}) &=\frac{\rho_i^{\rm{Dim 7}} (\rm{M^2},\rm{s_0})}{\rho_i (\rm{M^2},\rm{s_0})}\leq 5\%,
 \end{align}
 where $ \rho_i^{\text{Dim7}}( \rm{M^2}, \rm{s_0})$ represents the term with the highest dimension in the QCD formulation of $ \rho_i ( \rm{M^2}, \rm{s_0}) $.
 
\begin{widetext}
    \begin{figure}[htp]
\centering
  \includegraphics[width=0.44\textwidth]{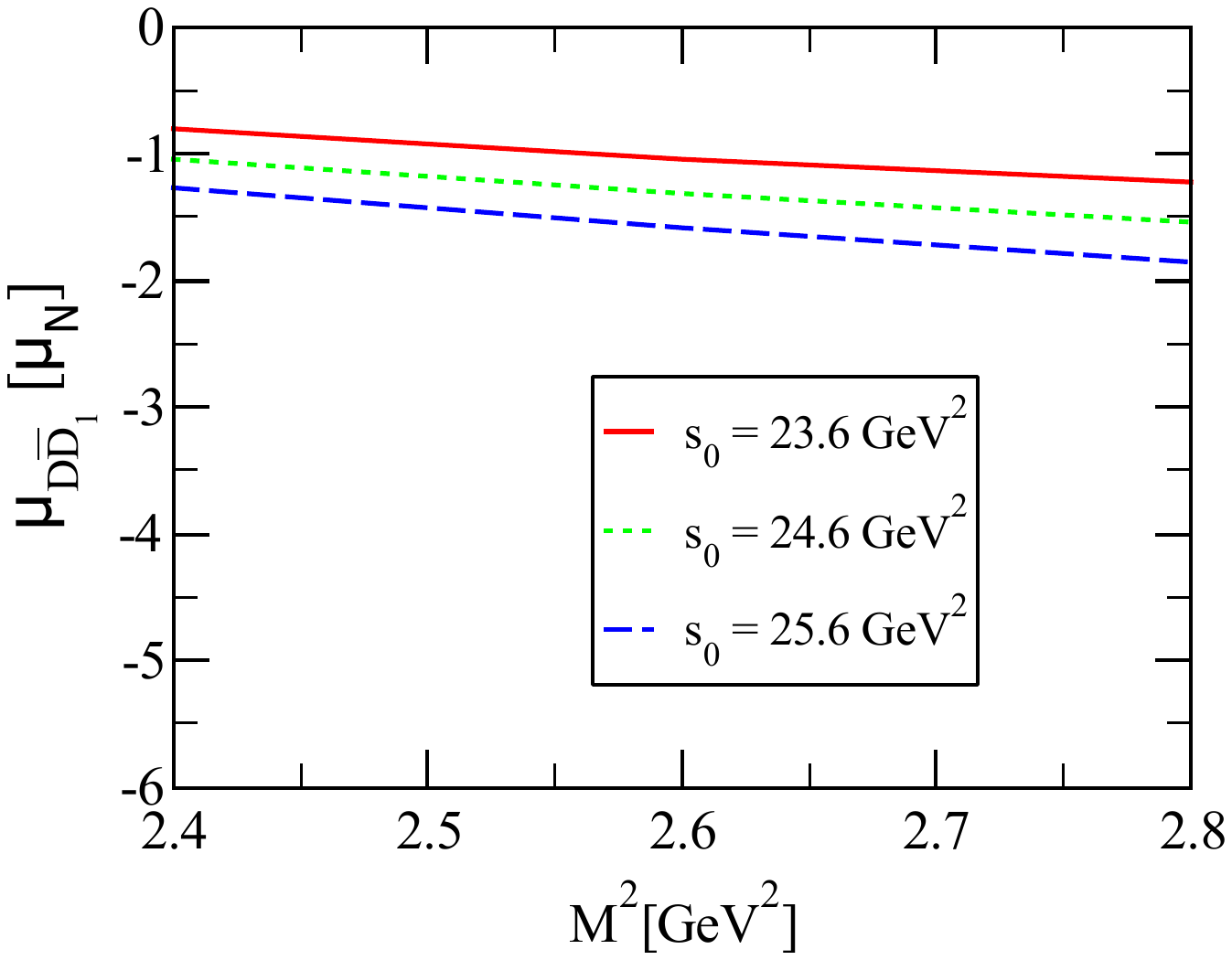} ~~~
    \includegraphics[width=0.44\textwidth]{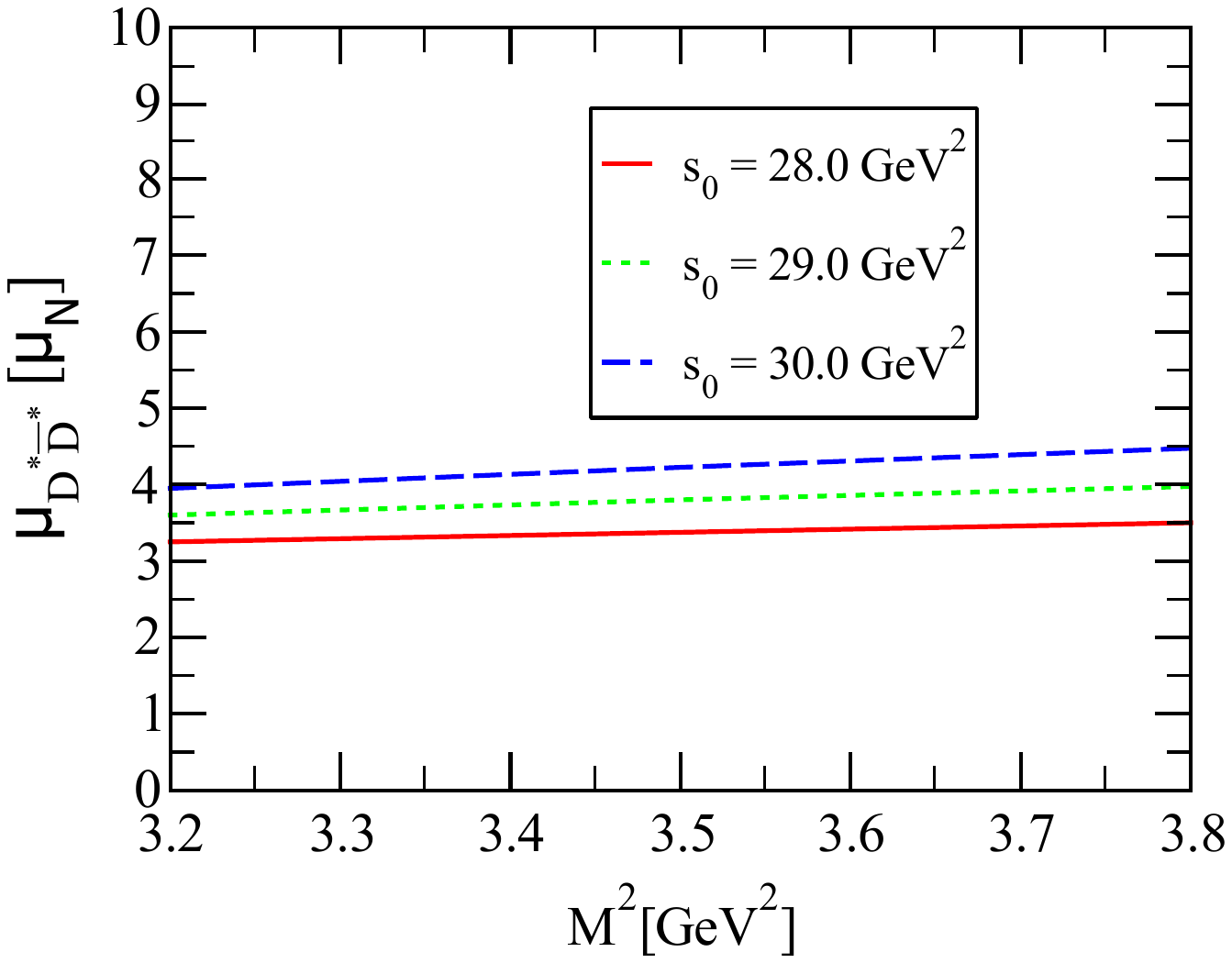}\\
     \includegraphics[width=0.44\textwidth]{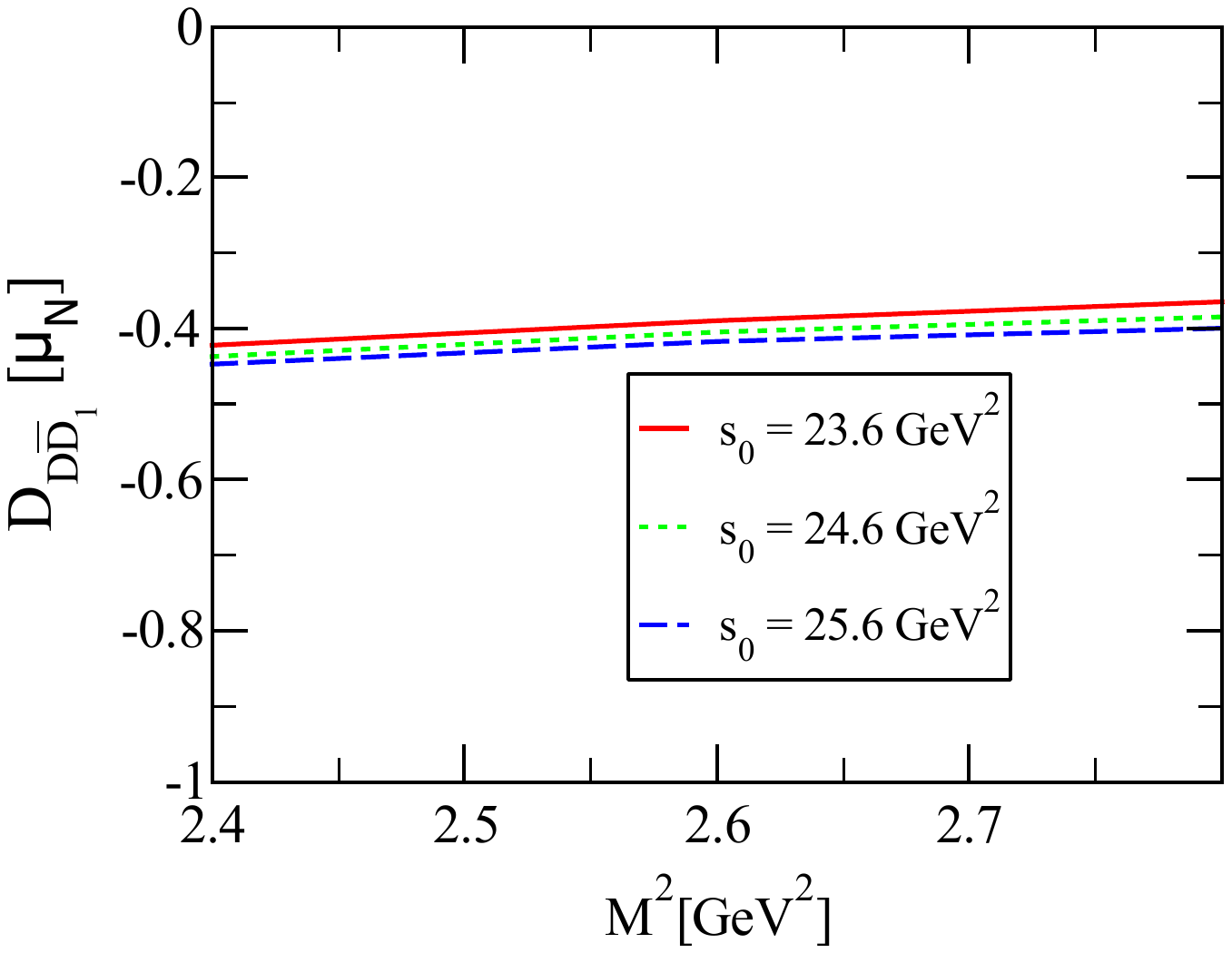} ~~~
    \includegraphics[width=0.44\textwidth]{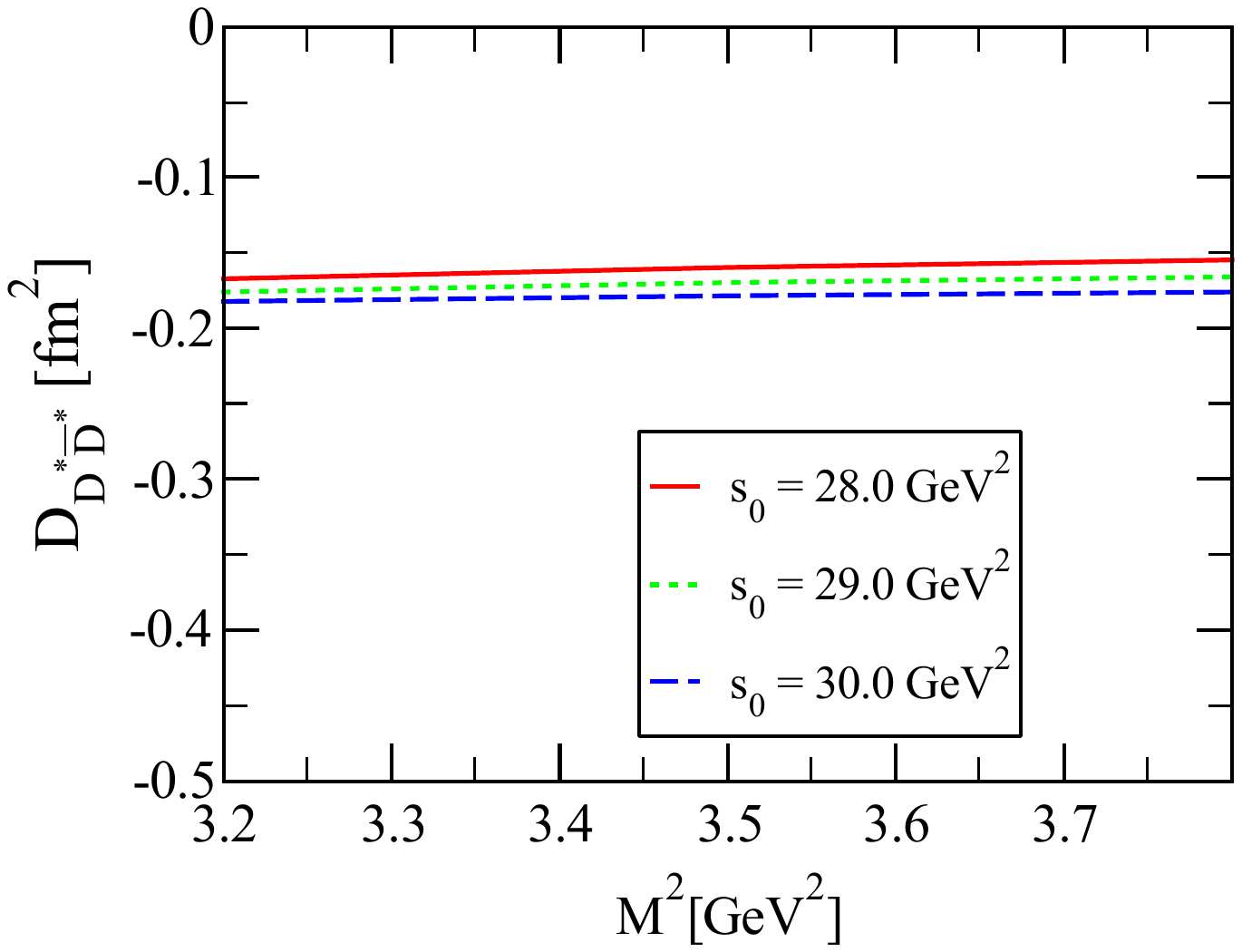}
  \caption{Magnetic and quadrupole ($\times 10^{-2}$) moments of the  $D\bar{D}_1$ and $D^*\bar{D}^*$ molecular states as a function of Borel mass at several continuum thresholds.}
 \label{figMsq}
  \end{figure}
  \end{widetext}

 Guided by the established requirements, the working regions of the auxiliary parameters are identified and presented in Table~\ref{table}. As shown in Table ~\ref{table}, the results satisfy all methodological requirements. Having successfully fulfilled these fundamental conditions of our approach, we now proceed with confidence in the reliability of our predictions. 
For a more thorough examination, Fig.~\ref{figMsq} presents the dependence of the calculated magnetic and quadrupole moments on variations of the auxiliary parameters. The illustration demonstrates a modest variation in the outcomes observed in these intervals, as anticipated. Notably, the results may exhibit some uncertainty due to the existence of residual dependencies. 
\begin{widetext}
  \begin{table}[htb!]
	\addtolength{\tabcolsep}{6pt}
	\caption{Working regions of $\rm{M^2}$ and $\rm{s_0}$
together with the PC and CVG for the electromagnetic properties of the molecular $D \bar D_1$ and  $D^* \bar D^*$   states.}
	\label{table}
	\begin{ruledtabular}
\begin{tabular}{lccccccc}
	   \\
	   Tetraquarks & $\rm{s_0}\,\,[\rm{GeV}^2]$&   $\rm{M^2}\,\,[\rm{GeV}^2]$& PC\,\,[$\%$] & CVG\,\,[$\%$]  & $\mu\, [\mu_N]$ &  $\mathcal{D}~[\mbox{fm}^{2}] (\times 10^{-2})$  \\
	   \\
	   \hline\hline
	  \\
$D \bar D_1$& [23.6, 25.6]&  [2.4, 2.8]& [56.42, 46.69]& $ \sim 2.0$ & $- 1. 41 \pm 0.50$ & $-0.40 \pm 0.10$ \\
\\
$D^* \bar D^*$& [28.0, 30.0]&  [3.2, 3.8]& [59.88, 45.20]& $\sim 1.0$ & $~~3.85 \pm 0.95$ & $-0.20 \pm 0.05$\\
\\
\end{tabular}
\end{ruledtabular}
\end{table}
 \end{widetext}

Through detailed numerical analysis, we have determined the magnetic and quadrupole moments for the $D \bar D_1$ and $D^* \bar D^*$ states. The presented results in Table \ref{table} incorporate full error propagation from all input parameters while systematically accounting for variations in auxiliary parameters. 
We would like to point out that the contributions to the total uncertainty are approximately as follows: 13\% from the mass of the tetraquarks, 22\% from their residues, 30\% from the threshold parameter $\rm{s_0}$, 7\% from the Borel mass $\rm{M^2}$, 11\% from the photon DAs, and the remaining 17\% from other parameters.  Our results suggest the following observations: 
  \begin{itemize}
\item The magnitude of the magnetic moments offers valuable information regarding the experimental accessibility of these physical observables. Given the significant order of magnitude of these values, we expect them to be accessible in future experimental investigations.
     
\item An analysis of the obtained results reveals that the dominant contribution arises from the short-distance interaction between quarks and the photon, corresponding to the perturbative component. In contrast, the remaining contributions stem from the long-distance interactions of light quarks with the photon, which are inherently non-perturbative in nature. The signs of the perturbative and non-perturbative contributions are found to be opposite. The perturbative contribution primarily determines the signs of both the magnetic and quadrupole moments.
  
\item The decomposition of the magnetic moment into its constituent quark contributions has been investigated, with the complete numerical results tabulated in Table~\ref{table3}. These analyses have been performed using the central values of the input parameters, with the uncertainties in these parameters not being considered. A deeper examination of the quark-by-quark contributions to the magnetic moments reveals that the predicted values are predominantly determined by light quarks, with nearly the entire contribution originating from these quarks. When a more comprehensive analysis is performed, the terms proportional to the c-quark are typically obtained with expressions scaling as $\sim e_c m_u$ and $\sim e_c m_d$. Since these terms yield extremely small contributions, the resulting c-quark contribution to the magnetic moment is found to be negligible.  When investigating the origin of this scaling, we find that in the OPE of the correlation function, the charm-quark magnetic moment contribution (\( e_c m_c \)) arises from diagrams where the photon couples perturbatively to the charm-quark line, as shown in Eqs.~(\ref{eq:QCDSide})-(\ref{eq:QCDSide1}), with the remaining light and heavy quark propagators taken fully into account. However, a more detailed analysis reveals that the \( e_c m_c \) terms undergo an almost exact cancellation, significantly suppressing their net contribution. Due to this interplay, the leading surviving terms instead scale as \( e_c m_u \) and \( e_c m_d \). This behavior is rooted in the particular Lorentz and Dirac structures of the interpolating currents and the OPE expansion, which suppress the contribution of the charm-quark magnetic moment due to cancellations among different operator structures.
%
  \begin{table}[htb!]
	\addtolength{\tabcolsep}{6pt}
	\caption{Flavor-decomposed magnetic and quadrupole ($\times 10^{-2}$) moments contributions for the $D\bar{D}_1$ and $D^*\bar{D}^*$ molecular states.}
	\label{table3}
	\begin{ruledtabular}
\begin{tabular}{lccccccc}
	   \\
	   Tetraquarks &   $\mu_{u} \,[\mu_N]$& $\mu_d \,[\mu_N]$& $\mu_{c} \,[\mu_N]$&$\mu_{tot} \,[\mu_N]$\\
	   \\
	   \hline\hline
	  \\
$D \bar D_1$&  $-0.950$& $-0.475$& $~~0.015$ &$-1.41$\\
\\
$D^* \bar D^*$&  $~~2.570$& $~~1.290$& $-0.010$&~~3.85\\
\\
\hline\hline 
\\
Tetraquarks &   $\mathcal D_{u} \,[\rm{fm}^2]$& $\mathcal D_d \,[\rm{fm}^2]$& $\mathcal D_{c} \,[\rm{fm}^2]$&$\mathcal D_{tot} \,[\rm{fm}^2]$\\
	   \\
	   \hline\hline
	  \\
$D \bar D_1$&  $-0.270$& $-0.130$& $0.00$ &$-0.40$\\
\\
$D^* \bar D^*$&  $-0.135$& $-0.065$& $0.00$&$-0.20$\\
\\
\end{tabular}
\end{ruledtabular}
\end{table}

\item A reverse correlation is observed between the contributions of light and heavy quarks. The signs of the resulting magnetic moments reflect the fundamental spin interactions among the constituent quarks. Specifically, the opposite signs of the magnetic moments associated with light and heavy quarks indicate that their spins are aligned in an anti-parallel manner within the molecular meson structures.

\item The calculated quadrupole moments of the $D\bar{D}_1$ and $D^*\bar{D}^*$ states yield non-zero values, suggesting a possible deformation in their charge distributions.  In the context of hadron structure, the sign of the quadrupole moment may offer insights into the geometric configuration, where a negative value ($\mathcal D < 0$) could imply oblate deformation, while a positive value ($\mathcal D > 0$) might suggest prolate deformation.  Our analysis yields quadrupole moment values with a negative sign, which could be interpreted as evidence for an oblate shape. An analysis of individual quark contributions to the quadrupole moments is additionally presented in Table~\ref{table3}. As can be observed from this table, the results appear to be entirely dominated by light-quark contributions. Our analysis suggests that the charm-quark contributions to the quadrupole moment may largely cancel out due to opposing sign contributions, resulting in a net negligible effect.

\item The mass of the $D\bar{D}_1$ molecular state is found to be consistent with the experimentally observed $Y(4360/4390)$ resonance. In Ref.~\cite{Ozdem:2022kck}, the magnetic moment of this state was examined within the framework of QCD light-cone sum rules, under the assumption of a compact tetraquark internal structure. 
In that study, they predicted the magnetic moment for this state to be:
$ \mu_{Y(4360/4390)} = 0.80^{+0.25}_{-0.21}~\mu_N.$ The result obtained can be seen to differ significantly from the one in this study. The observed discrepancy suggests that the magnetic moment could serve as a sensitive probe of hadronic internal structure, potentially distinguishing between molecular and compact tetraquark configurations.

\item Before concluding, it is worthwhile to briefly address the experimental challenges associated with determining the magnetic moments of the \( D\bar{D}_1 \) and \( D^*\bar{D}^* \) molecular states. Due to their extremely short lifetimes, conventional methods—such as spin precession measurements or vector meson–electron scattering—are not practically feasible for these systems, making direct experimental determination of their magnetic properties highly challenging. The magnetic moments of short-lived hadrons can instead be extracted indirectly through a multi-step process. First, the hadron of interest is produced in a high-energy collision. Subsequently, it emits a soft photon, which effectively acts as a probe in the presence of an external magnetic field. Finally, the hadron decays, and its properties are reconstructed from the decay products to infer the magnetic moment.  An alternative approach involves studying the electromagnetic properties of vector mesons through their radiative production and decay. Theoretical studies suggest that the energy and angular distributions of emitted photons provide a viable means of determining these properties~\cite{LopezCastro:1997dg}. This technique has been employed to extract the magnetic moment of the $\rho$ meson, using preliminary data from the BaBar Collaboration for the $e^+ e^- \rightarrow \pi^+ \pi^- 2 \pi^0$ process, within the center of mass energy range from $0.9$ to $2.2$ GeV. The extracted value for the magnetic moment of the $\rho$ meson was found to be $\mu_\rho = 2.1 \pm 0.5~e/{2 m_\rho}$ units~\cite{GarciaGudino:2013alv}.   An alternative approach is also available for indirectly determining the electromagnetic properties of vector mesons. The primary motivation behind this method is based on the soft photon emission from hadrons, as proposed in Ref.~\cite{Zakharov:1968fb}, where a technique for determining the electromagnetic multipole moments is outlined. The core concept of this technique is that the photon encodes information regarding the magnetic moment, as well as higher-order multipole moments, of the particle from which it is emitted.  The matrix element for the radiative process can be formulated by the following relation as a function of the photon energy, \(E_\gamma\).
\begin{align}
 M \sim A \, (E_\gamma )^{-1}+ B\,(E_\gamma )^0 + \mbox{higher-order multipole moments}. 
\end{align}
In this expression, \((E_\gamma)^{-1}\) and \((E_\gamma)^0\) denote the contributions from the electric charge and the magnetic moment, respectively. 
Thus, the magnetic moment of the states being analyzed can be extracted by measuring the cross-section or decay width of the radiative process, while disregarding the small contributions from the linear and higher-order terms in \(E_\gamma\). The magnetic moment of the $ \Delta^+(1232)$ particle has been obtained by applying the steps proposed by this method~\cite{Pascalutsa:2004je, Pascalutsa:2005vq, Pascalutsa:2007wb,   Kotulla:2002cg, Drechsel:2001qu, Machavariani:1999fr, Drechsel:2000um, Chiang:2004pw, Machavariani:2005vn}.  Although direct experimental measurements are currently not feasible for short-lived hadrons, their electromagnetic properties can be inferred through careful analysis of radiative processes with the help of various indirect approaches.

We now explore specific radiative decay channels and experimental processes that may serve as sensitive probes of the electromagnetic structure of the $Y$ states.  The $Y(4360/4390)$ resonance can be produced experimentally via the initial-state radiation (ISR) mechanism in high-energy $e^+e^-$ collisions. In this process, one of the incoming leptons emits a photon before annihilation, effectively reducing the center-of-mass energy of the $e^+e^-$ system to match the mass of the $Y(4360/4390)$. The ISR production channel is given by
\[
e^+ e^- \rightarrow \gamma_{\mathrm{ISR}} + Y(4360/4390),
\]
and has been successfully used by experiments such as Belle and Belle II to study vector charmoniumlike states. 
Once produced, the $Y(4360/4390)$ can undergo radiative decays into lower charmonium states, such as
\[
Y(4360/4390) \rightarrow \gamma\, \psi(2S),
\]
followed by the decay of $\psi(2S)$ into lepton pairs or via $\psi(2S) \rightarrow \pi^+ \pi^- J/\psi$ with $J/\psi \rightarrow \ell^+ \ell^-$, which are clean experimental signatures. 
This decay channel is particularly interesting because its rate and photon energy distribution are sensitive to the internal electromagnetic structure of the $Y(4360/4390)$. If the $Y(4360/4390)$ is a compact charmonium state, this transition is expected to occur with a relatively large branching fraction. On the other hand, if the $Y(4360/4390)$ has a hadronic molecular structure---such as a $\bar{D} D_1$ bound state---the overlap with $\psi(2S)$ may be suppressed, leading to a reduced transition amplitude. Therefore, studying this radiative decay at experiments like Belle II or BESIII may provide critical insights into the underlying structure of the $Y$ states.

Another possible class of radiative transitions involves the decay of the $Y(4360/4390)$ into the $\chi_{cJ}$ (with $J=0,1,2$) charmonium states through the emission of a photon:
\begin{equation}
Y(4360/4390) \to \gamma \, \chi_{cJ}.
\end{equation}
These transitions correspond to electric dipole (E1) processes, as they involve a change in the orbital angular momentum of the system ($\Delta L = 1$), connecting an $S$-wave initial state to $P$-wave final states. In the conventional charmonium interpretation, such transitions are generally expected and have been studied extensively in bottomonium and charmonium systems. However, if the $Y(4360/4390)$ is instead a hadronic molecular state, such as a $\bar{D} D_1$ or $\bar{D}^* D^*$ bound system, the coupling to compact $c \bar{c}$ final states like $\chi_{cJ}$ could be significantly suppressed due to the mismatch in internal structure and wave-function overlap. As a result, the observation (or suppression) of these E1 transitions can serve as an indirect probe of the underlying nature of the $Y$ states. Experimentally, such radiative transitions can be searched for in high-luminosity electron-positron colliders like Belle II or BESIII via processes such as
\begin{equation}
e^+ e^- \to Y(4360/4390) \to \gamma \, \chi_{cJ} \to \gamma \, (J/\psi \, \gamma),
\end{equation}
where the final state $J/\psi \to \ell^+ \ell^-$ provides a clean experimental signature. The photon energy and angular distributions in these processes may carry imprints of the internal electromagnetic structure of the $Y$ resonance.

The $Y(4360/4390)$ resonance can also be studied through its possible radiative decays into open-charm final states involving a photon, such as
\begin{equation}
Y(4360/4390) \rightarrow D^{(*)} \bar{D}^{(*)} \gamma.
\end{equation}
These decay modes are particularly relevant if the $Y(4360/4390)$ is interpreted as a hadronic molecular state, composed of a weakly bound pair of heavy mesons such as $\bar{D} D_1$ or $\bar{D}^* D^*$. In this scenario, the photon emission can proceed through the radiative decay of one of the constituent mesons, for instance:
\[
Y(4360/4390) \to D^* \bar{D} \to D \bar{D} \gamma,
\]
or
\[
Y(4360/4390) \to D^* \bar{D}^* \to D \bar{D} \gamma.
\]

Such decays are expected to be more favorable in a molecular configuration due to the natural presence of $D^*$ mesons, which can decay via well-known electromagnetic transitions like $D^* \to D \gamma$. The soft photon emitted in this process carries information about the long-range electromagnetic dynamics between the constituents, which is sensitive to the spatial extension of the molecular wave-function. Experimentally, the $Y(4360/4390)$ can be produced via the ISR mechanism in high-energy $e^+ e^-$ collisions:
\begin{equation}
e^+ e^- \to \gamma_{\mathrm{ISR}} + Y(4360/4390).
\end{equation}
Once produced, the $Y(4360/4390)$ may decay radiatively into open-charm final states, and the detection of soft photons in association with $D$ mesons provides a distinctive experimental signature. The measurement of photon energy spectra and angular correlations in such processes can reveal key features of the $Y(4360/4390)$'s internal structure. If the $Y(4360/4390)$ is a compact charmonium state, such transitions are expected to be highly suppressed due to the absence of open-charm components in the wave-function. In contrast, in the molecular picture, such decays are enhanced, and their observation would offer strong evidence for the extended spatial structure and non-conventional nature of the resonance. Therefore, studying the $Y(4360/4390) \to D^{(*)} \bar{D}^{(*)} \gamma$ processes in experiments like Belle II or BESIII can serve as a crucial test to distinguish between compact and molecular interpretations of the $Y(4360/4390)$.

While the proposed radiative channels are theoretically promising, we acknowledge the experimental challenges. Soft photons in \( D^{(*)} \bar{D}^{(*)} \gamma \) decays require advanced calorimetry and low-energy triggers. Small branching fractions may necessitate high-statistics datasets. 
Non-resonant backgrounds can be suppressed through kinematic fitting, for example by imposing the condition \( M(D^{(*)} \bar{D}^{(*)}) \sim m_Y \). Moreover, quantitative estimates of transition rates in molecular models are still limited; future theoretical work could compute \( Y \to \gamma \psi(2S) \) widths using overlap integrals within effective frameworks (e.g., within HQET or potential models).

\end{itemize}

\section{Discussion and Outlook}\label{sum}

As in previous decades, a comprehensive understanding of the intricate internal configuration of hadrons continues to be a central objective within both experimental and theoretical hadron physics. This pursuit plays a pivotal role in advancing our knowledge of QCD and critically evaluating the robustness and accuracy of the theoretical models developed to date. Furthermore, deciphering the underlying mechanisms of exotic states, both those currently observed and those anticipated in future experiments, remains a pressing and unresolved challenge. Motivated by this, in the present study, we investigate the electromagnetic properties of vector molecular states within the framework of the QCD light-cone sum rule approach. Assuming a molecular structure composed of meson pairs, we derive sum rules for the relevant form factors by incorporating photon distribution amplitudes up to twist-4 accuracy. From these form factors, we extract static electromagnetic observables, such as the magnetic dipole and quadrupole moments.  
Studying the electromagnetic properties of the $D \bar D_1(2420)$ and $D^* \bar D^*(2400)$ molecular states yields valuable insights into their internal structure and geometric deformation. The calculated magnetic and quadrupole moments reveal distinct signatures that differentiate these states from compact tetraquark configurations, highlighting the sensitivity of electromagnetic observables to the underlying hadronic dynamics. The dominance of light-quark contributions to the magnetic moments and the oblate charge distribution suggested by the quadrupole moments offer a deeper understanding of the spatial arrangement and spin interactions within these exotic hadrons.

The results of this study align with the broader goal of deciphering the complex nature of multiquark states, which remain a frontier in hadron physics. The observed discrepancies between the magnetic moments of molecular and compact tetraquark configurations, such as the Y(4360/4390) resonance, underscore the potential of electromagnetic properties as diagnostic tools for distinguishing between different structural models. This capability is particularly crucial given the ongoing experimental efforts to identify and characterize exotic hadrons at facilities like LHCb, Belle, and BESIII.

\bibliography{Vectormolecule.bib}
\bibliographystyle{elsarticle-num}

\end{document}